# On the enhanced cosmic-ray ionization rate in the diffuse cloud towards ζ Persei

**Gargi Shaw[1], G. J. Ferland[1], R. Srianand[2], N. P. Abel[1], P. A. M. van Hoof[3], & P. C. Stancil[4]**

## Abstract

The spatial distribution of the cosmic-ray flux is important in understanding the Interstellar Medium (ISM) of the Galaxy. This distribution can be analyzed by studying different molecular species along different sight lines whose abundances are sensitive to the cosmic-ray ionization rate. Recently several groups have reported an enhanced cosmic-ray ionization rate ($\zeta = \chi_{CR}\zeta_{standard}$) in diffuse clouds compared to the standard value, $\zeta_{standard}$ (=$2.5\times10^{-17}$ $s^{-1}$), measured toward dense molecular clouds. In an earlier work we reported an enhancement $\chi_{CR}$ =20 towards HD185418. McCall et al. have reported $\chi_{CR}$ =48 towards ζ Persei based on the observed abundance of $H_3^+$ while Le Petit et al. found $\chi_{CR} \approx 10$ to be consistent with their models for this same sight line. Here we revisit ζ Persei and perform a detailed calculation using a self-consistent treatment of the hydrogen chemistry, grain physics, energy and ionization balance, and excitation physics. We show that the value of $\chi_{CR}$ deduced from the $H_3^+$ column density, $N(H_3^+)$, in the diffuse region of the sightline depends strongly on the properties of the grains because they remove free electrons and change the hydrogen chemistry. The observations are largely consistent with $\chi_{CR} \approx 40$, with several diagnostics indicating higher values. This underscores the importance of a full treatment of grain physics in studies of interstellar chemistry.

---

[1] University of Kentucky, Department of Physics and Astronomy, Lexington, KY 40506; gshaw@pa.uky.edu, gary@pa.uky.edu, npabel2@uky.edu

[2] IUCAA, Post bag 4, Ganeshkhind, Pune 411007, India; anand@iucaa.ernet.in

[3] Royal Observatory of Belgium, Ringlaan 3, 1180 Brussels, Belgium; p.vanhoof@oma.be

[4] University of Georgia, Department of Physics and Astronomy and Center for Simulational Physics, Athens, GA 30602; stancil@physast.uga.edu

## 1. Introduction

$\zeta$ Persei is a type B star (Snow 1977, Perryman et al. 1997) situated at a distance of 300 parsecs (RA: 03h58m57.93s, Dec: +35$^0$47′28″, J2000). The diffuse cloud towards $\zeta$ Persei is important because of the measurement of a large $H_3^+$ column density, $N(H_3^+)$. $H_3^+$ plays an important role in the ion-molecule chemistry in the Interstellar Medium (ISM). It is created mainly via the gas-phase reaction $H_2 + H_2^+ \rightarrow H_3^+ + H$ once $H_2^+$ is produced by cosmic-ray ionization of $H_2$. Therefore, a large $N(H_3^+)$ suggests the presence of an enhanced cosmic-ray ionization rate along this sightline (McCall et al. 2003, hereafter MC03). A nice chronological discussion of the problem can be found in Le Petit et al. (2004; hereafter LP04).

Several groups have proposed that an enhanced cosmic–ray ionization rate, $\chi_{CR}$(= $\zeta/\zeta_{standard}$, where $\zeta_{standard}$ is the standard rate), may vary across the galaxy (e.g., Federman et al. 1996). MC03 have derived an enhanced $\chi_{CR}$ = 48 towards $\zeta$ Persei by combining an experimental determination of the dissociative recombination rate of $H_3^+$ ($k_e$) with the observed $N(H_3^+)$ and assuming a hydrogen number density of 250 cm$^{-3}$. On the other hand, LP04 modeled the diffuse cloud as a two-phase plane parallel slab that is illuminated on one side by twice the galactic background radiation field. They found that $\chi_{CR}$ =10 can reproduce most column densities within a factor of 3, but they required shocks to reproduce the observed $N(CH^+)$ and the populations of excited rotational levels of $H_2$ ($J$= 2, 3, 4 and 5). Hence they concluded that enhancement of $\chi_{CR}$ is less than that required by MC03.

In an earlier paper (Shaw et al. 2006, hereafter S06), we showed when the particle density is well constrained by fine-structure level populations, the observed ortho-para temperature and high-$J$ excitations of $H_2$ are very sensitive to $\chi_{CR}$. Further, it was found in S06 that $\chi_{CR}$ =20 reproduces the observed temperature and high-$J$ excitation in the diffuse interstellar medium (ISM) towards HD185418 without requiring additional shocked gas.

Performing an independent analysis is one way of verifying the high $\chi_{CR}$ requirement towards $\zeta$ Persei. In a diffuse cloud, $H_3^+$ is mainly destroyed by dissociative recombination with electrons ($H_3^+ + e^- \rightarrow H + H + H$ or $H_2 + H$). Assuming steady state,

$$\zeta_{\rm standard}\chi_{CR} = k_e n(H_3^+) n_e \ / n(H_2)\,, \qquad (1)$$

where $n(H_2)$, $n(H_3^+)$, and $n_e$ are densities of $H_2$, $H_3^+$, and $e^-$, respectively. For a given $k_e$ and $n(H_2)$, the relationship between $N(H_3^+)$ and $\chi_{CR}$ depends on $n_e$. To determine all of the parameters necessary to evaluate $\chi_{CR}$ , we perform a detailed calculation using a self-consistent treatment of the hydrogen chemistry, grain physics, energy and ionization balance, and excitation physics. We find that grains are very important in determining the free electron density, $n_e$. The grains along this sight line have a small ratio of total to selective extinction $R_V$, indicating a grain size distribution with an enhanced number of small particles. The small grains are very effective at absorbing free electrons, changing the ionization balance of the cloud. This alters the $\chi_{CR}$ deduced from the observed $N(H_3^+)$ as illustrated below.

# 2. Description of the Model and Parameters

## *2.1. Chemical and thermal state*

In this work, we used version 07.02 of the spectral simulation code Cloudy, which was last described by Ferland et al. (1998). It performs a self-consistent calculation of the thermal, ionization, and chemical balance of both the gas and dust. The chemical network consists of ~$10^3$ reactions with 71 species involving H, He, C, N, O, Si, S, and Cl (Abel et al. 2005). The effects of ion charge transfer on grain surfaces are described in Abel et al. (2005). The comparison with other existing codes (Roellig et al. 2007) finds good agreement in the PDR (photon-dominated region) limit.

Further, we include silicate and graphite size-resolved grains and determine the grain charge and photoelectric heating self-consistently (van Hoof et al. 2004). The calculations use ten size bins each for silicate and graphite. We adjust the size distribution to reproduce the observed extinction, as discussed below. We do not include polycyclic aromatic hydrocarbons (PAHs) due to a lack of observational evidence along this line of sight.

Our treatment of $H_2$ is described in Shaw et al. (2005, hereafter S05). The $H_2$ chemistry network consists of various state-specific formation and destruction processes. The model explicitly includes cosmic-ray excitation to the singlet states and triplet state via secondary electrons generated by cosmic rays (Dalgarno et al. 1999). Here we improve this model by adopting, for the singlet states B, C, B', and D, the Born approximation cross section

$$\sigma = A_{ul} \frac{m\lambda^3}{2\pi a_0 h} \frac{1}{k^2} \ln\left(\frac{k'+k}{k'-k}\right), \qquad (2)$$

(Liu & Dalgarno 1994) in terms of the dipole-allowed electronic spontaneous emission transition probabilities. *k* and *k '* are the incoming and outgoing momenta of the electrons, and all other symbols have their usual meaning. We assume an average incoming energy of 20 eV. The excitation rate is estimated by rescaling this cross section in terms of hydrogen ionization cross section (Shemansky et al. 1985) and then rescaling with the cosmic-ray excitation rate of Lyα. The cross section for the dissociative triplet state b is taken from Dalgarno et al. (1999, Fig. 4b), rescaled in terms of the hydrogen ionization cross section (Shemansky et al. 1985), and then multiplyed with the cosmic-ray excitation rate of Lyα. These processes excite the electronic states eventually populating the higher vib-rotational levels of the ground state, thus changing the populations of levels within *X*.

In addition to the chemistry described in Abel et al. (2005), we now consider condensation of CO, $H_2O$, and OH onto grain surfaces following the formalism of Hasegawa, Herbst, & Leung (1992), Hasegawa & Herbst (1993), and Bergin et al. (1995). For each molecule, condensation is balanced with desorption due to thermal and cosmic ray evaporation. Grain surface reactions between molecules are not considered. Condensation is important when the gas and dust temperature fall below 20-25 K (Bergin et al. 1995). Removing a molecule from the gas phase will alter many of the reaction rates and will cause the molecular abundances to change from the case of pure gas-phase chemistry. Although included, it was found that molecules do not condensate on grain surfaces in the environments considered here as the grain temperature is higher than 25 K.

Bound-electron Compton scattering is the most efficient photoionization process for photon energies above ~10 keV (Osterbrock & Ferland 2006, Fig 11.5). We assume a cross section that is scaled from the Klein-Nishina electron cross section multiplied by the number of valence electrons. This is included as a general ionization and heating process for all molecules and for molecules condensed onto grain surfaces.

The gas kinetic temperature is determined at each point in the cloud including all relevant heating and cooling processes as detailed in Abel et al. (2005). In a typical calculation, the temperature will range from ~5000 K, where a significant amount of $H^+$ is present, to ~50 K in well-shielded molecular regions. The cloud is assumed to be sufficiently old for the chemical, ionization, and thermal balances to have reached time-steady conditions.

### *2.2. Focus of this paper*

Gas is present in three distinct phases (ionized, diffuse, and dense) along this sight line (Snow 1977, van Dishoeck & Black 1986; LP04). Most of the column density is in a low-density PDR-like medium containing $H_2$ and $C^+$. Thin layers that are predominantly molecular or ionized are also present. Our emphasis is on using $N(H_3^+)$ to derive $\chi_{CR}$, while simultaneously accounting for other observables that are produced in the same environment where $H_3^+$ forms. The abundance of $H_3^+$ is proportional to the $H_2$ abundance and inversely proportional to $n_e$, so we concentrate on the diffuse phase, where most of the $H_2$ resides and investigate different processes that influence $n_e$.

The observed $N(C_2)$ and $N$(CN) are produced in a thin dense molecular phase (Gredel et al. 2002). The total column density of this phase is only $N(H_2)$ ~$10^{19}$ $cm^{-2}$ (LP04) and should, therefore, give only a small contribution to the total $N(H_3^+)$. In what follows, we model the diffuse phase and do not attempt to model the conditions in the dense molecular phase. We therefore make no attempt to reproduce the $N(C_2)$ and $N$(CN) column densities and will not further develop the LP04 model of the dense phase..

The presence of a significant column density of $Si^{+2}$ and $S^{+2}$ indicates that an $H^+$ region is also present along this sight line. As described below, the radiation field includes a hydrogen-ionizing continuum (Black 1987), which produces a thin layer of $H^+$ on the surface of the diffuse cloud. Nearby hot stars contribute additional hydrogen-ionizing photons. These ions are attributed to an $H^+$ region near the background star as in S06 and no attempt is made to reproduce them here.

### *2.3. Cloud geometry & illuminating radiation*

We consider a plane-parallel geometry with radiation striking both sides. van Dishoeck & Black (1986) found that this geometry reproduced the physical conditions within typical diffuse interstellar clouds. We also used this geometry to reproduce the properties of a well studied sightline towards HD185418 (S06). The Galactic background radiation field given by Black (1987) and the cosmic rays are the only source of photoelectric heating and ionization in the calculations. We interpolate an FUV continuum between the UV and X-ray observations (**13.6 to 54 eV**) and our interpolation of Black (1987) is shown in Table 1a. We parameterize the intensity of the incident radiation by $\chi$, the ratio of the assumed incident radiation field to the Galactic background. The scale factor multiplies the entire continuum, except for the CMB (Cosmic Microwave Background) which is added separately to the rescaled ISM field. The radiative transport of the full continuum, from radio through X-rays, is done including pumping by the incident continuum, atomic and molecular lines, and background opacities. It also includes line overlap by the hundreds of thousands of electronic transitions that determine the destruction and excitation of $H_2$ (S05) and other lines.

### *2.4. Model parameters*

We have adopted $R_v = 2.85$ (Cardelli et al. 1989) and $E(B\text{-}V) = 0.32$ giving $A_v/N(\mathrm{H_{tot}}) = 5.7\times10^{-22}$ cm$^2$. where $N(\mathrm{H_{tot}})$ is the column density of H in all forms. The grain-size distribution was modified to approximate this extinction law. Following Mathis et al. (1977) a power-law size distribution of grains was adopted, $dn/da \propto a^{-3.5}$, where $a$ is the radius of the grain and d$n$ is the number of grains with a radius lying within $a$ and $a+da$. The number of smaller grains is larger than in a standard mixture. However, most of the grain mass resides in the larger grains as the volume is $\propto a^3$. We consider an astronomical silicate and graphite with 10 size bins ranging between $a_{\min} = 0.001\mu$m and $a_{\max} = 0.2$ μm. The Mathis et al. model of the standard ISM used $a_{\min} = 0.005\mu$m and $a_{\max} = 0.25$ μm. The charges and temperatures of the grains are determined self-consistently (van Hoof et al. 2004). We use the size-dependent grain temperature to find the $H_2$ formation rate using Cazaux & Tielens (2002). In section 3.3.4 we show how the deduced $\chi_{CR}$ depends on the assumed grain properties.

The width of absorption lines is important for continuum fluorescence and shielding. It is determined by both thermal and turbulent motions. Turbulence sets the linewidth since thermal motions are small for low temperature and heavy particles. The observed Doppler line width $b$ is 2.5 km s$^{-1}$ for the $H_2$ lines (Snow 1977; LP04), which we adopt. This may not actually be a physical microturbulence but could include macroscopic motions of different clouds. As a test, we also ran a model without turbulence but keeping all the other parameters identical to our final model derived below. The predicted column densities remain the same except for CO, which increased by ~0.2 dex. The total $H_2$ column density is not modified by turbulence because most of the $N(H_2)$ resides in lower levels and electronic lines from lower levels are on the damped portion of the curve of growth.

We concentrate on the low-density phase since that accounts for most of the observed $H_3^+$ column density. We assume $N(H_{tot}) = 10^{21.2}$ cm$^{-2}$ and that the total hydrogen density is constant across the cloud, as would be the case where magnetic (Heiles & Crutcher 2005) or turbulent pressure (Tielens & Hollenbach 1985) dominates the gas equation of state.

The density is determined by matching the observed column densities of C I, C I$^*$, and C I$^{**}$ ($^3P_0$, $^3P_1$, $^3P_2$) with the computed ionization, chemical, and thermal balance. Note that the excitation rates and densities of various colliders vary across the cloud since they are obtained self-consistently. We find a total hydrogen density of $n(H_{tot})$ = 80 cm$^{-3}$. The Galactic starlight background enhancement $\chi$ is obtained by fitting $S^0/S^+$, $Mg^0/Mg^+$, $Fe^0/Fe^+$, $C^0/C^+$ and $Ca^0/Ca^+$. We find $\chi$ close to 2.

The resulting parameters are listed in Table 1b and are reasonably close to those found by LP04. This best fitting model will be described further below.

The remaining free parameter is the cosmic-ray ionization rate enhancement $\chi_{CR}$. This is obtained by matching to the column densities of CO, $H^0$, $H_2$ in various *J* levels, NH, and $H_3^+$. In additional calculations, all the parameters listed in Table 1b are keep fixed, and $\chi_{CR}$ is varied to illustrate its effect on the predicted column densities. We find that $S^0/S^+$, $Mg^0/Mg^+$, $Fe^0/Fe^+$, $C^0/C^+$ and $Ca^0/Ca^+$ have a small dependence on $\chi_{CR}$ in addition to their dependence on $\chi$. These ionization ratios change by less than 0.2 dex for $1 < \chi_{CR} < 100$, while the ratios change nearly linearly with $\chi$. The $2\sigma$ errors in the observed values are larger than 0.2 dex so we did not use them to constrain $\chi_{CR}$.

## 3. Results and Discussion

### *3.1 A model with $\chi_{CR} = 1$*

As a first test, we consider $\chi_{CR} = 1$ and compute the hydrogen-ionization structure as a function of cloud depth. We assume the density and radiation field derived in the previous section. Figure 1 shows half of the symmetric cloud. We stopped our calculation at half of the observed value of $N(H_{tot}) = 10^{21.2}$ cm$^{-2}$ since the cloud is assumed to be symmetric. At the illuminated face the gas is moderately ionized, with $n(H^0) \sim 35$ cm$^{-3}$ and $n(H^+) \sim 45$ cm$^{-3}$. In deeper regions hydrogen is molecular with the $H_2$ fraction $f(H_2) \sim 0.99$. The electron density $n_e$ and electron temperature $T_e$ are displayed as functions of cloud depth in Figure 2. The density $n_e$ varies from 45 to 0.014 cm$^{-3}$ while the temperature ranges from 5370 to 38 K across the cloud. We see that $H^+$ is the main electron donor at the illuminated face. The predicted column densities are listed in the third column of Table 2. This model clearly underpredicts $N(H^0)$, $N(C^0)$, $N$(NH), $N$(CO), $N$(OH), $N(H_3^+)$, and the column densities of the $J \geq 2$ rotational levels of $H_2$.

## *3.2Models with varying $\chi_{CR}$*

Cosmic-rays heat and ionize the gas. Increasing $\chi_{CR}$ increases the densities of $e^-$, $H^0$ and $H^+$ in gas that is predominantly $H_2$. Further, $H^0$ and $H^+$ undergo exchange collisions with $H_2$ and induce ortho-para conversions. The upper half of Figure 3 shows how the column densities of various rotational levels of $H_2$ change with varying $\chi_{CR}$. This clearly illustrates that populations in the higher rotational levels of $H_2$ depend on $\chi_{CR}$. The lower half of Figure 3 displays the column densities of CO, OH, $H_3^+$, and NH. $\chi_{CR} \geq 40$ reproduces the observed column densities of the rotational levels of $H_2$ ($J$ = 0, 1, 2, 3, 4, 5), $H^0$, CO, $H_3^+$, and NH, within a factor of 2 or better and reproduces the $H_2$ ortho-para temperature. However, the predicted column density of OH is too large for $\chi_{CR} \geq 10$. Interestingly, the CO and OH column densities change together as $\chi_{CR}$ changes, suggesting that the problem is actually in the CO/OH balance. The dashed vertical line highlights $\chi_{CR} = 40$, the best fit value.

## *3.3 Characteristics of the best-fit model*

### 3.3.1 Overview of the results

Black lines in Figure 1 plot the hydrogen-ionization structure with $n(H^0) \sim 35$ cm$^{-3}$ and $n(H^+) \sim 45$ cm$^{-3}$ at the illuminated face for the $\chi_{CR} = 40$ model. These number densities at the illuminated face resemble the earlier model with $\chi_{CR} = 1$ since the shallow regions are photon-dominated. Deep inside the cloud hydrogen is mainly in $H_2$ and $f(H_2)$ is ~0.78. The value of $f(H_2)$ is smaller than the earlier model with $\chi_{CR} = 1$ due to enhanced cosmic-ray destruction of $H_2$. The temperature also increases deep inside the cloud compared to the $\chi_{CR}$ = 1 model. Similarly, black lines in Figure 2 show that $n_e$ and $T_e$ vary from 45 to 0.034 $cm^{-3}$ and 5370 to 51 K as functions of cloud depth and clearly shows that the enhanced $\chi_{CR}$ increases $T_e$ and $n_e$ deep inside the cloud. The fifth column of Table 2 lists the predicted column densities with $\chi_{CR}$ = 40.

The merit of this model is that the enhanced $\chi_{CR}$ not only reproduces the column densities of $H_3^+$, $H^0$, CO, and NH within a factor of 2 but also the column densities of higher-rotational levels of $H_2$ and the measured $H_2$ ortho-para temperature. In our calculation we use the dissociative recombination rate of $H_3^+$ measured by MC03. For a comparison, we also show the resulting $H_3^+$ abundance with a dissociative recombination rate of $H_3^+$ as derived by Datz et al. (1995) and tabulated by Stancil et al. (1998). The Datz et al. (1995) rate, being larger than that reported by MC03, produces a lower $H_3^+$ column density. Note that we do not require shocks to reproduce the column densities of the higher-rotational levels of $H_2$. It is achieved by the cosmic-ray excitation of the ground state levels. Higher $\chi_{CR}$ excites the electronic states more efficiently and enhances the populations of the higher vib-rotational levels of the ground state. Earlier van Dishoeck & Black (1986) also reproduced the column densities of the higher rotational levels of $H_2$ towards ζ Persei without shocks by using an assumed radial gradient in temperature and density.

Our model reproduces the column densities of $H^0$, $O^0$, $C^0$, $C^+$, $S^0$, $S^+$, $Si^+$, $H_3^+$, and CO, NH within a factor of 2. We reproduce the observed ratio of $S^0/S^+$, $Mg^0/Mg^+$, $Fe^0/Fe^+$, $C^0/C^+$, $Ca^0/Ca^+$ and the corresponding column densities within 2σ of the observational uncertainties which suggests that our predicted $\chi/n_e$ is consistent with the observed data. Furthermore, the column densities determined from [C I] fine-structure excitations are reproduced verifying our predicted gas density.

### 3.3.2 CH and $CH^+$

Our calculation, like all those that use a standard ISM chemistry network, underpredicts the abundances of CH and $CH^+$. The basic problem is that the reactions forming these molecules have substantial energy barriers and do not occur at low temperatures. Investigations have suggested that these species can be produced by either shocks or non-thermal chemistry. Zsargó & Federman (2003) favor non-thermal chemistry. In the non-thermal chemistry driven by MHD (magneto-hydrodynamical) waves, an effective temperature is defined based on the turbulence, which then allows the above reactions to occur. Shocks are another alternative. In this case the gas temperature is high and the reactions can occur.

If $CH^+$ is produced by shocks, OH will be produced as well. At high temperatures, $CH^+$ is formed by $C^+ + H_2 \rightarrow CH^+ + H$ (reaction I) and OH by $O + H_2 \rightarrow OH + H$ (reaction II). There is evidence suggesting that the formation of $CH^+$ is not always associated with the production of OH (for example, towards $\xi$ Per). Federman (1996) showed that reaction (I) can occur without (II) if MHD waves only interact with the ions and this chemistry scheme can reproduce observed $N(CH^+)$ without producing excessive $N$(OH). We have added the non-thermal chemistry caused by MHD waves as an option in Cloudy. We also included the ability to couple the MHD waves to all reactions, or to just the reactions where one of the reactants is an ion. Our predicted column densities of CO, OH, CH, $CH^+$, CN, $C_2$, $C_3$ with non-thermal chemistry are listed in table 2 (the last column marked with "c"). We find reasonable agreement with the observed $N(CH^+)$ with little increase in $N$(OH) for non-thermal chemistry.

### 3.3.3 Other species

The CO/OH ratio is almost independent of $\chi_{CR}$ (CO/OH ~ 1.07 for $\chi_{CR}$ >20, see Figure 3). Our predicted CO/OH ratio is ~10 times smaller than the observed value suggesting that the problem is in the chemistry rather than $\chi_{CR}$. In the ISM, the formation of OH starts with the creation of $O^+$ followed by ion-molecule reactions. CO is formed mainly via $C^+ + OH \rightarrow CO + H^+$. The rate coefficient adopted for this process-by Cloudy is taken from Dubernet et al. (1992). CO can also form from $HCO^+$. If we accept these values of the CO formation rate coefficients, then the CO/OH ratio depends primarily on the photodissociation rates of CO and OH. Our CO and OH photodissociation rates are taken from van Dishoeck (1988) and the UMIST database respectively. While there has been considerable effort in the past to obtain reliable CO and OH photodissociation rates, some uncertainty may still remain. Recent studies have found that experimental oscillator strengths for photodissociation of CO are larger than those included in UMIST (Federman et al. 2001, 2003**;** Eidelsberg et al. 2006). This creates a faster dissociation rate when the gas is optically thin to the CO electronic lines, and more self-shielding when the lines become optically thick. Experimental oscillator strengths are not available for the majority of the CO electronic lines, establishing an uncertainty in the predictions. Improved radiative rates for CO along with a detailed treatment of the radiative transfer using more modern grain physics are required. In the case of OH, the photodissociation cross sections are highly accurate, but standard, and potentially, out-of-date ISM grain models were used to deduce attenuated photo-destruction rates. Further, these rates were fitted to simple functions which may culminate in considerable uncertainty, up to nearly a dex. For instance, a factor of 3 increase in the photodissociation rate of OH and a factor of 3/2 decrease in the photodissociation rate of CO results in column densities of OH (log $N$ = 14.0 $cm^{-2}$) and CO (log $N$ = 14.66 $cm^{-2}$) that are within a factor of two of the observed range. If these uncertainties are considered the differences are not serious.

As mentioned above, we do not try to reproduce the molecules that form in a dense molecular region ($C_2$, $C_3$ and CN) or the highly ionized species that form in the $H^+$ region near the background star ($S^{+2}$ and $Si^{+2}$). We do list our predictions of these species for the diffuse phase.

For comparison we list our predicted column densities for $\chi_{CR}$ = 10 in the fourth column of Table 2. Our predicted OH column density matches the observation better for $\chi_{CR}$ = 10 but the predicted column densities of the other species ($H^0$, $O^0$, $C^0$, $H_3^+$, CO, NH, and higher rotational levels of $H_2$) are lower than the observed values. In contrast, we reproduce the observed column densities within a factor of 2 for $\chi_{CR}$ = 40 with the one exception of OH. We cannot decrease $\chi_{CR}$ in order to reproduce OH because in that case the model will underpredict other species, including the high-$J$ excitation in $H_2$, by more than a factor of 3. Keeping in mind the possible uncertainties in the photodissociation rates of OH and CO, the $\chi_{CR}$ ~40 case is certainly a possible solution in the diffuse cloud towards ζ Persei.

### 3.3.4 The grain-size distribution

Figure 4 and Table 3 show how details of the grain physics affects the $H_3^+$ column density and so the deduced $\chi_{CR}$. We perform tests with different grain-size distributions, keeping all other parameters fixed. The figure shows the ratio $n(H_3^+)/n(H_2)$, which is proportional to $\chi_{CR} / n_e$ (see eqn. 1). This sight line is characterized by $R_v$ = 2.85 and we use a custom size distribution with a greater than standard number of small particles to approximate the extinction on this line of sight. This will increase the rate of charge exchange with ions, with small grains capturing free electrons and absorbing FUV radiation. Since $\chi_{CR}$ is held fixed, the changes shown in Figure 4 are mainly due to changes in $n_e$. Had we assumed, like many

ISM studies (for example MC03), that the electron density is equal to the carbon density, we would have needed a value of $\chi_{CR}$ almost two times larger to reproduce $N(H_3^+)/N(H_2)$. To see the effect of collisional grain physics for this sight line we turned off grain-ion charge exchange as well as electron capture by grains. As a result $n_e$ increases, mainly because elements with low ionization thresholds stay ionized and donate more electrons to the gas. This results in a lower value of $N(H_3^+)/N(H_2)$ . We find that grain-ion charge exchange plays an important role in this model.

The temperature is derived from the balance between heating and cooling processes. The gas heating occurs mainly via grain photo-electric heating, cosmic-ray heating, heating due to $H_2$ dissociation and collisional de-excitation, whereas the gas cooling occurs mainly via collisionally excited fine-structure atomic and molecular lines. In table 2 we list the excitation temperature ($T_{10}$) derived from the $J$=1 and $J$=0 levels of $H_2$. Deep inside the cloud, cosmic-rays also heat the gas and an increase in $\chi_{CR}$ will increase the temperature (and hence increase $T_{10}$). For $\chi_{CR}$ =40, the cosmic-ray heating rate deep inside the cloud is ~ $9.3\times10^{-25}$ erg cm$^{-3}$s$^{-1}$ which is roughly half of the total heating.

S06 noticed that $\chi_{CR}$ = 20 increased $T_{10}$ by a factor of two above $\chi_{CR}$ = 1 along the line of sight towards HD185418. However, towards ζ Persei the temperature did not increase significantly even though we have a higher $\chi_{CR}$ = 40. One contributor is the smaller $R_v$ = 2.85 and hence the presence of more small grains towards ζ Persei. For a given mass, the smaller grains have more surface area and produce more heating compared to larger grains. As a result, deep inside the cloud the heating due to cosmic rays is a smaller fraction of the total. Furthermore, the density along the ζ Persei line of sight is higher. Since the effects of cosmic ray heating are proportional to $\chi_{CR}/n_H$, the larger $\chi_{CR}$ is compensated by the larger $n_H$.

### 3.3.5 Unobserved species

We list the predicted column densities for some unobserved species in Table 4. The predicted column densities of $OH^+$ and $H_2^+$ for $\chi_{CR}$ = 40 are ~ 4 times larger than that for $\chi_{CR}$ = 10. Future detection of these species will help to distinguish between these two cases. It should be possible to measure hydrogen recombination lines along this sight line but these would form in the $H^+$ region that produces $S^{+2}$ and $Si^{+2}$ rather than in cosmic ray ionized gas. In our model high-$J$ levels of $H_2$ are excited by cosmic rays. If instead these excitations are produced in shocks then one will be able to detect the differences in the velocity of the absorption line using high resolution spectroscopic data. Absence of any such velocity difference will support the models presented here.

Recently Padoan & Scalo (2005) have shown that self-generated MHD waves produce an enhanced cosmic-ray density in diffuse clouds compared to those found in dense clouds. They have shown the variation of the cosmic-ray density as a function of $n_H$ in their figure 1 and suggest a scaling law like $\chi_{CR} \propto n_i^{0.5}$, where $n_i$ is the ion density. The density of the diffuse cloud towards ζ Persei lies near the top of the curve predicting $\chi_{CR}$. In an earlier paper (S06) we required 20 times $\chi_{CR}$ to reproduce the observed column density along a sight line for which $n_H = 27$ cm$^{-3}$. Assuming $\chi_{CR} \propto n_H^{0.5}$ and rescaling the $\chi_{CR}$ found by S06 by the ratio of densities according to Padoan & Scalo's prescription we should get ~34.5 times $\chi_{CR}$ in the diffuse cloud towards ζ Persei which lies very close to the lower limit of our range of suggested $\chi_{CR}$.

# 4 Conclusions

The observation of a large $H_3^+$ column density towards ζ Persei raises a controversy about the value of cosmic-ray ionization rate towards this line of sight. Earlier McCall et al. (2003) have derived an enhanced $\chi_{CR} = 48$ towards ζ Persei by assuming $n(H) = 250$ cm$^{-3}$ and that most of the electron density $n_e$ comes from carbon photoionization. On the other hand, LP04 found that $\chi_{CR} = 10$ can reproduce almost all observed column densities within a factor of 3 based on a detailed model consisting of two thermal phases plus three shocks. Whereas, our detailed model reproduces many of the observed column densities in the diffuse cloud towards ζ Persei with the help of an enhanced cosmic-ray ionization rate. We find that with the exception of OH, an enhancement of $\chi_{CR} \geq 40$ can reproduce the observed column densities of $H_3^+$, $H^0$, CO, NH and various $H_2$ lines, within a factor of 2. The merit of our model is that we do not require shocks to reproduce the observed $N(CH^+)$ and populations of high rotational levels of $H_2$ ($J$= 2, 3, 4 and 5). We use non-thermal chemistry driven by MHD waves which has an advantage of producing the observed $N(CH^+)$ without producing excessive OH. The high rotational levels of $H_2$ are produced as a result of excitation by cosmic-rays.

The value of the cosmic-ray ionization rate is model dependent since some assumptions come into play. Tests presented above show that the grain physics has a significant effect on the deduced $\chi_{CR}$. Small grains, which are overabundant along this low-$R_v$ sight line, neutralize ions, remove free electrons, and extinguish FUV radiation. These effects by themselves change the deduced $\chi_{CR}$ by more than a factor of two. This raises the question of the composition of the small grains – this is uncertain with some favoring PAHs rather than graphite, as we assume. The main unknown however is the grain size distribution itself which introduces a further factor of 2 uncertainty in addition to the unknowns presented by the chemical network and cloud geometry.

# 5 Acknowledgement

We thank F. Le Petit and B. J. McCall for their valuable comments. GS and NPA acknowledge support from the Computational Center for Sciences (CCS) of the University of Kentucky. GJF thanks the NSF, NASA, and STScI for support. RS and GJF acknowledge support from DST/INT/US (NSF-RP0-115)/2002. PvH acknowledges support from the Belgian Science Policy Office under the IAP P5/36 grant scheme. GS and GJF acknowledge the hospitality of IUCAA. PCS acknowledges support from NSF grant AST0607733 and NASA grant NNG05GD81G.

# 6 References

Abel, N. P., Ferland, G. J., Shaw, G., & van Hoof, P. A. M. 2005, ApJS, 161, 65

Bergin, E. A., Langer, W. D., & Goldsmith, P. F. 1995, ApJ, 441, 222

Black, J. H. 1987, Interstellar processes, ed. Hollenbach, D. J., Thronson, H. A. Jr., (Dordrecht: Reidel) 731

Black, J. H., & van Dishoeck, E.F. 1987, ApJ, 322, 412

Bohlin, R. C. 1975, ApJ, 200, 402

Cardelli, J. A., Clayton, G. C., & Mathis, J. S. 1989, ApJ, 345, 245

Cardelli, J. A., Meyer, D. M., Jura, M., & Savage, B. D. 1996, ApJ, 467, 334

Cartledge, S. I. B., Meyer, D. M., Lauroesch, J. T., & Sofia, U. J. 2001 ApJ, 562, 394

Cazaux, S., & Tielens, A. G. G. M  2002, ApJ, 575,29

Chaffee, F. 1974, ApJ, 189, 427

Chaffee, F., & Lutz, B. L. 1977, ApJ, 213, 394

Dalgarno, A., Yan, M., & Liu, W. 1999, ApJS, 125, 237

Datz, S., Sundström, G., Biedermann, Ch., Broström, L., Danared, H., Mannervik, S.,

Dubernet, M. L., Gargaud, M., & McCarroll, R. 1992, A&A, 259, 373

Eidelsberg, M., Sheffer, Y., Federman, S. R., Lemaire, J.L., Fillion, J.H., Rostas, F., & Ruiz, J. Astro-ph/0605186.

Federman, S. R. 1982, ApJ, 257, 125

Federman, S. R., Fritts, M., Cheng, S., Menningen, K. M., Knauth, D. C., & Fluk, K. 2001, ApJs, 134, 133

Federman, S. R., Weber, J., & Lambert, D. L. 1996, ApJ, 463, 181

Felenbok, P., & Roueff, E. 1996, ApJ, 465, 57

Ferland, G. J., Korista, K. T., Verner, D. A., Ferguson, J. W., Kingdon, J. B., & Verner, E. M. 1998, PASP, 110, 761

Gredel, R., Pineau des Forêts, G., & Federman, S. R. 2002, A&A, 389, 993

Jenkins, E. B., & Shaya, E. J. 1979, ApJ, 231, 55

Hasegawa, T. I., & Herbst, E. 1993, MNRAS, 261, 83

Hasegawa, T. I., Herbst, E., & Leung, C. M. 1992, ApJS, 82, 167

Heiles, C., & Crutcher, R. 2005, LNP, 664, 137

Hobbs, L. M. 1969, ApJ, 157, 135

Hobbs, L. M. 1974, ApJL, 188, 67

Jura, M., & Meyer, D. M. 1985, ApJ, 294, 238

Le Petit, F., Roueff, E., & Herbst, E. 2004, A&A, 417, 993 (LP04)

Liu, W., & Dalgarno, A. 1994, ApJ, 428, 769L

Maier, J. P., Lakin, N. M., Walker, G. A. H., & Bohlender, D. A. 2001, ApJ, 553, 267

Mathis, J. S., Rumpl, W., & Nordsieck, K. H. 1977, ApJ, 217, 425

McCall, B. J., Huneycutt, A. J., Saykally, R. J., Geballe, T. R., Djuric, N., Dunn, G. H., Semaniak, J., Novotny, O., Al-Khalili, A., Ehlerding, A., et al. 2003, Nature, 422, 500 (MC03)

Meyer, D. M., & Jura, M. 1985, ApJ, 297, 119

Meyer, D. M., & Roth, K. C. 1991, ApJ, 376, 49

Osterbrock, D.E. & Ferland, G.J. 2006, *Astrophysics of Gaseous Nebulae and Active Galactic Nuclei*, 2nd Edition (University Science Books)

Padoan, P., & Scalo, J. 2005 ApJL, 624, 97

Perryman, M. A. C., Lindegren, L., Kovalevsky, J. et al 1997, A&A, 323, 49

Roellig, M. et al. astro-ph/0702231

Shaw, G., Ferland, G. J., Abel, N. P., Stancil, P. C., & van Hoof, P. A. M. 2005, ApJ, 624, 794 (S05)

Shaw, G., Ferland, G. J., Srianand, R., & Abel, N. P. 2006 ApJ, 639, 941(S06)

Shemansky, D.E., Ajello, J.M., & Hall, D. T. 1985, ApJ, 296,765

Snow, T. P. 1977, ApJ, 216, 724

Stancil, P. C., Lepp, S., & Dalgarno, A. 1998, ApJ, 509, 1

Tielens, A. G. G. M., & Hollenbach, D. 1985, ApJ, 291, 722

van Hoof, P.A.M., Weingartner, J.C., Martin, P.G., Volk, K., & Ferland, G.J. 2004, MNRAS, 350, 1330

van Dishoeck, E. F., & Black, J. H. 1986, ApJS, 62, 109

van Dishoeck, E. F., & Black, J. H. 1988, ApJ, 334, 771

van Dishoeck, E. F., & Black, J. H. 1989, ApJ, 340, 273

Welty, D. E., Hobbs & L. M., York, D. G. 1991, ApJS, 75, 425

White, R. E. 1973, ApJ, 183, 81

Zsargó, T., & Federman, S. R.  2003, ApJ, 589, 319

# 7 Figures

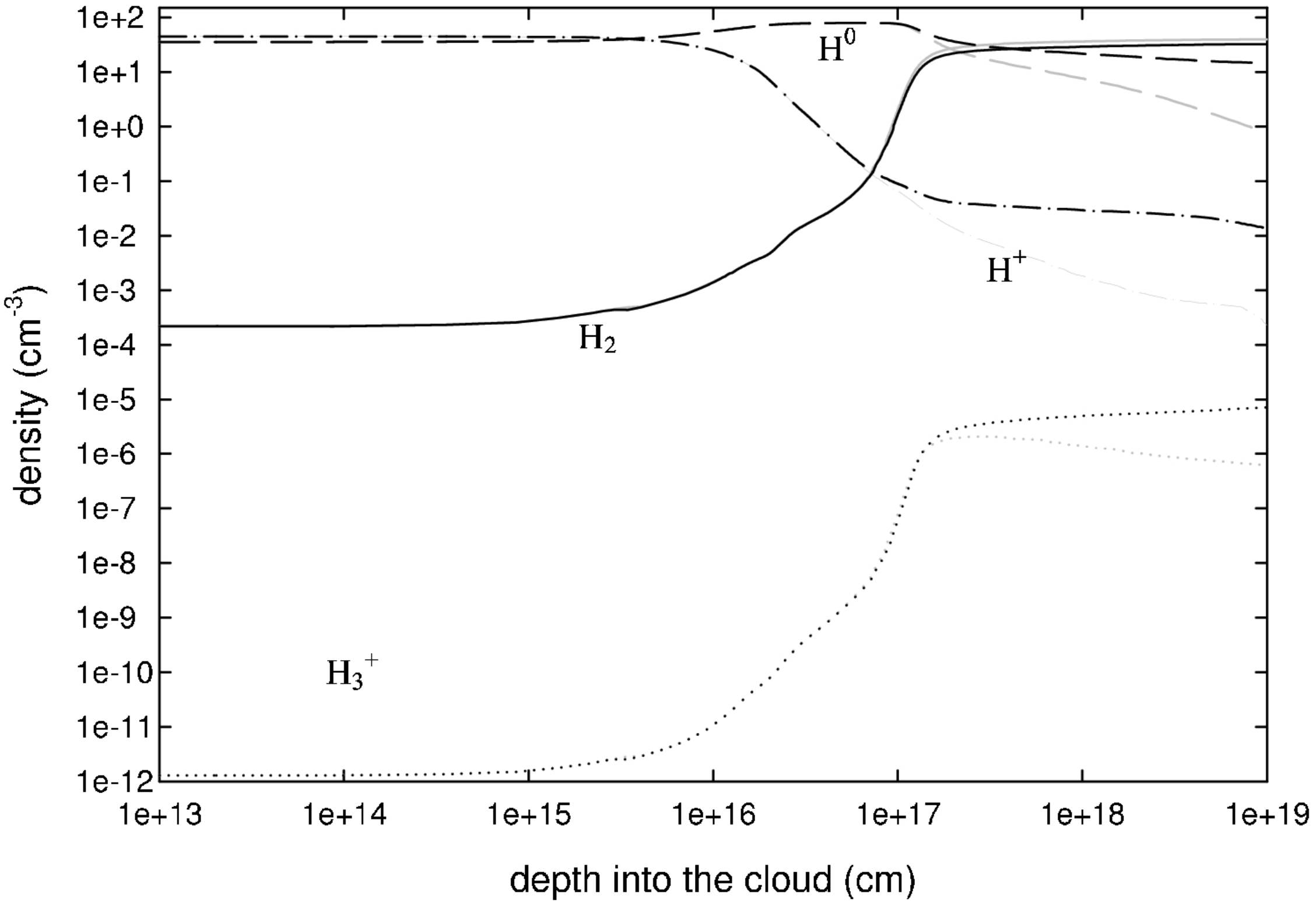

Figure 1: Hydrogen ionization structure as a function of cloud depth. The gray and the black lines represent cases with $\chi_{CR} = 1$, $\chi = 2$, $n_H = 80$ cm$^{-3}$ and $\chi_{CR} = 40$, $\chi = 2$, $n_H = 80$ cm$^{-3}$ respectively. This plot represents one half of an assumed symmetrical cloud. Due to numerical resolution the first derivatives of all quantities are not zero at the mid point.

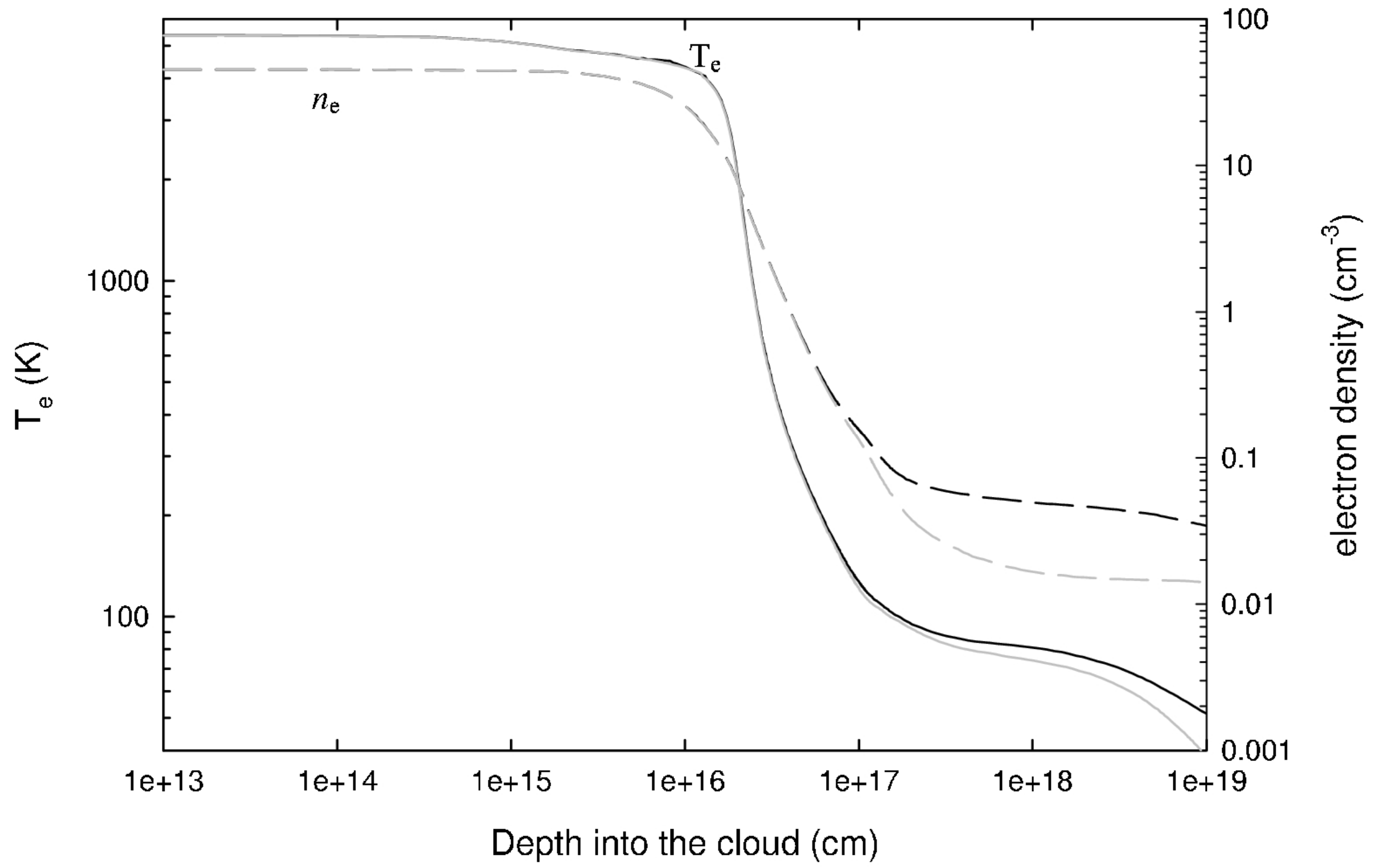

Figure 2: Variation of electron density and electron temperature as a function of cloud depth to the center. The gray and the black lines represent cases with $\chi_{CR}$ = 1, $\chi$ = 2, $n_H$ = 80 cm$^{-3}$ and $\chi_{CR}$ = 40, $\chi$ = 2, $n_H$ = 80 cm$^{-3}$ respectively.

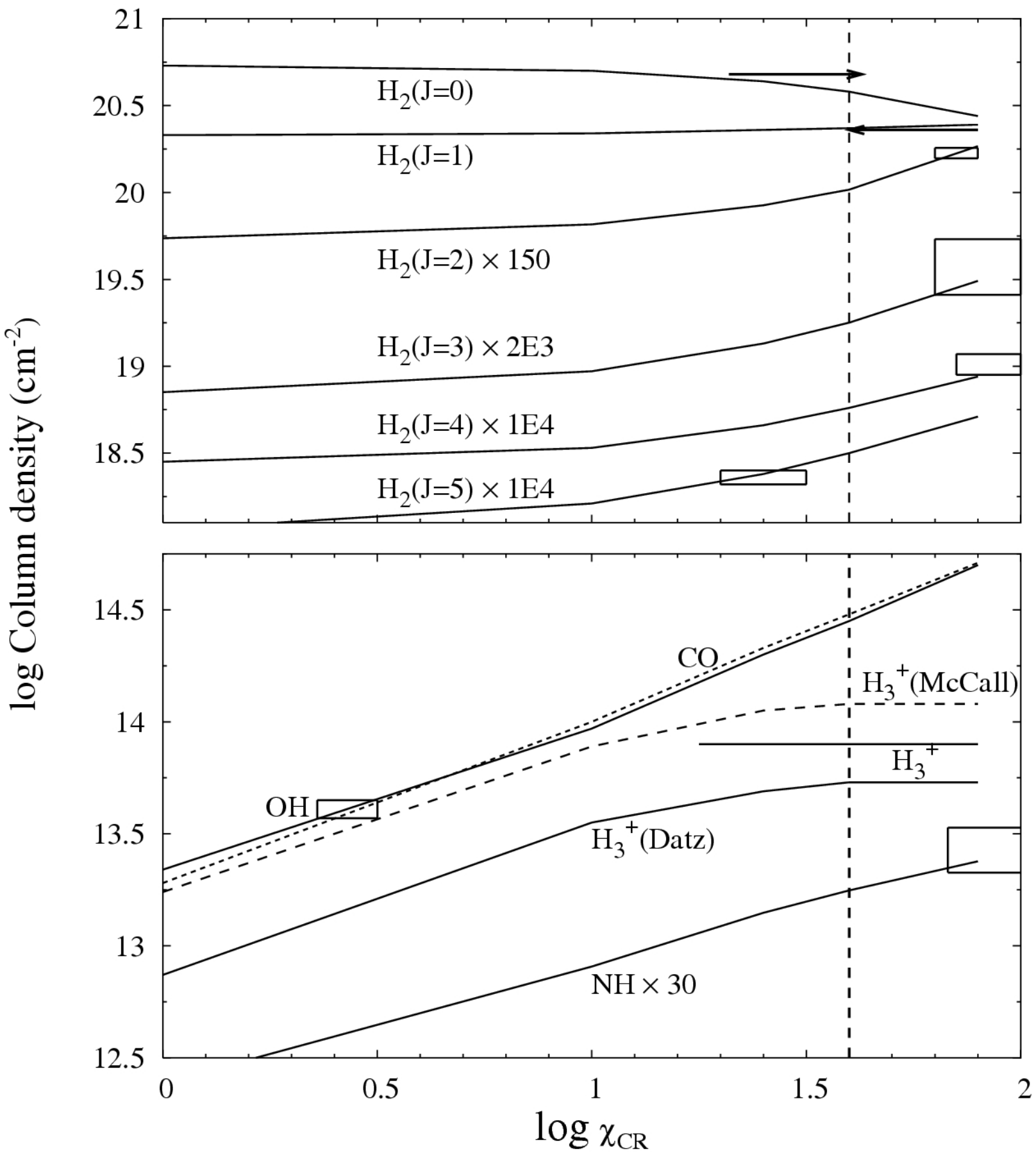

Figure 3: Variation of column densities as a function of $\chi_{CR}$. The boxes represent the range of observed values. NH, $H_2$(*J*=2), $H_2$(*J*=3), $H_2$(*J*=4), and $H_2$(*J*=5) column densities are multiplied by 30, 150, $2\times10^3$, $10^4$, and $10^4$ respectively to fit the range of the plot. The predicted $H_3^+$ column density is shown with two different dissociative recombination rate of $H_3^+$. The observed value of CO lies beyond the scale of this plot. The horizontal line shows the observed column density of $H_3^+$.

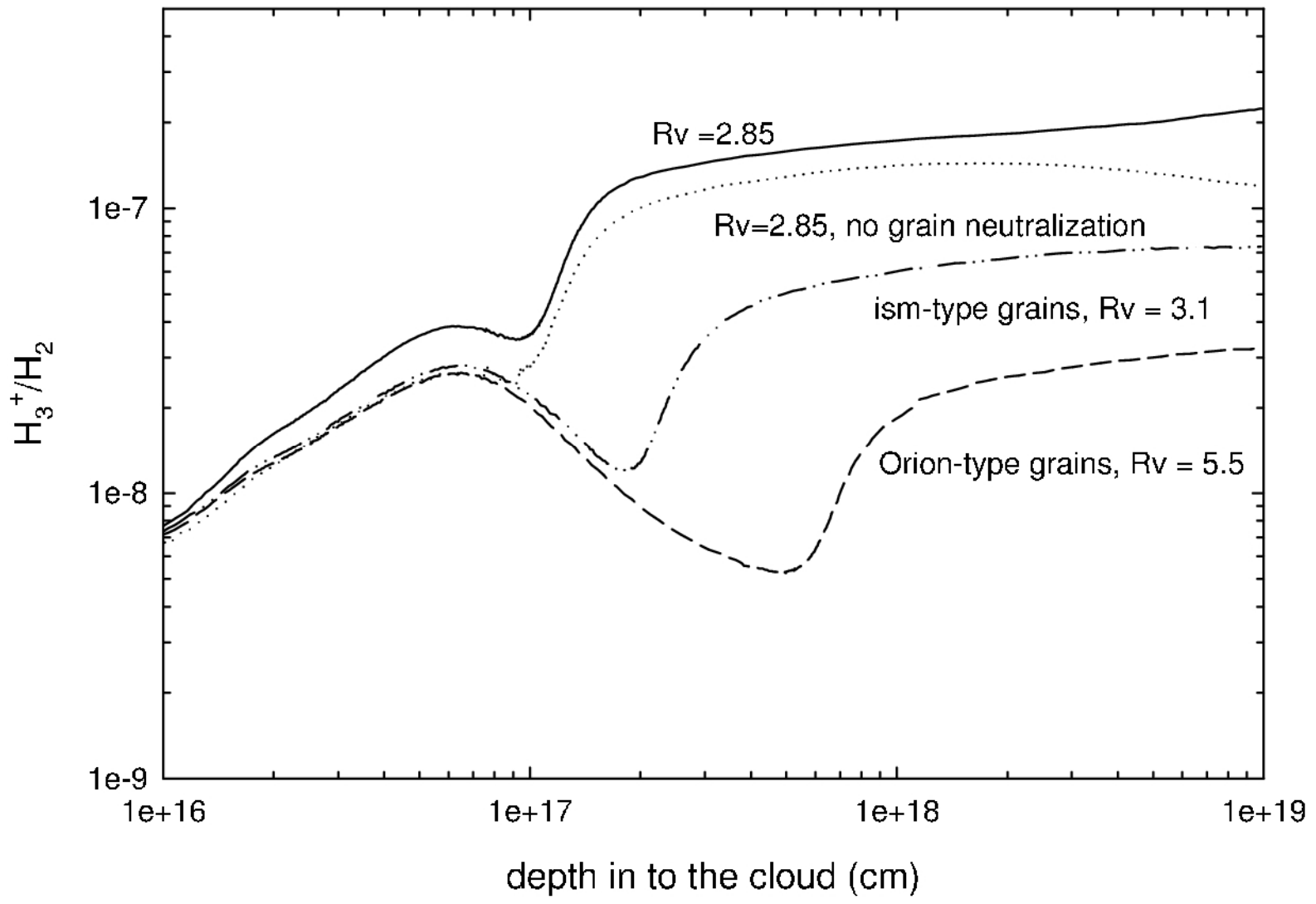

Figure 4: The variation of $n(H_3^+)/n(H_2)$, which is proportional to $\chi_{CR}/n_e$, as a function of cloud depth, for four different assumptions concerning the grain physics. Three grain size distributions are shown, and, for the ζ Per case, results both with and without the capture of electrons onto grains and grain-ion charge exchange are shown.

# 8 Tables

**Table 1a: ISM radiation field**

| log ν (Hz) | log ν $f_\nu$ (erg $cm^{-2}$ $s^{-1}$) | log ν (Hz) | log ν $f_\nu$(erg $cm^{-2}$ $s^{-1}$) |
|---|---|---|---|
| 6.00 | -16.708 | 14.14 | -2.30 |
| 10.72 | -2.96 | 14.38 | -1.79 |
| 11.00 | -2.47 | 14.63 | -1.79 |
| 11.23 | -2.09 | 14.93 | -2.34 |
| 11.47 | -2.11 | 15.08 | -2.72 |
| 11.55 | -2.34 | 15.36 | -2.55 |
| 11.85 | -3.66 | 15.54 | -2.62 |
| 12.26 | -2.72 | 16.25 | -5.68 |
| 12.54 | -2.45 | 17.09 | -6.45 |
| 12.71 | -2.57 | 18.00 | -6.30 |
| 13.10 | -3.85 | 23.00 | -11.30 |
| 13.64 | -3.34 | | |

**Table 1b: Model Parameters**

| parameters | values |
|---|---|
| $n_H$($cm^{-3}$) | 80 |
| $\chi$ | 2 |
| log C/H | -3.9 |
| log O/H | -3.5 |
| log S/H | -4.9 |
| log Si/H | -4.5 |
| log Ca/H | -8.8 |
| log Cl/H | -7.5 |
| log Mg/H | -5.9 |
| log Fe/H | -7.0 |
| log N/H | -4.2 |
| Turbulence | 2.5 km $s^{-1}$ |

**Table 2: Column densities of various species**

| Column densities (derived from references in the parentheses) | Observed log $N$ ($cm^{-2}$) | Predicted log $N$ ($cm^{-2}$) | | |
|---|---|---|---|---|
| | | $\chi_{CR} = 1$ | $\chi_{CR} = 10$ | $\chi_{CR} = 40$ |
| $H^0$ (1) | 20.76 to 20.85 | 19.90 | 20.20 | 20.55 |
| $C^0$ (2) | 15.46 to 15.56 | 15.22 | 15.36 | 15.60 |
| C I (2) | 15.44 to 15.52 | 15.13 | 15.27 | 15.49 |
| C I* (17) | 14.11 to 14.36 | 14.40 | 14.57 | 14.89 |
| C I** (2) | 14.04 to 14.14 | 13.74 | 13.81 | 14.15 |
| $Ca^0$ (19) | 9.70 to 9.82 | 9.37 | 9.54 | 9.83 |
| $S^0$ (21) | 13.12 to 13.52 | 13.20 | 13.34 | 13.59 |
| $O^0$ (20) | 17.71to 17.72 | 17.69 | 17.69 | 17.69 |
| $Ar^0$ (2) | 14.65 to 15.01 | 15.20 | 15.20 | 15.20 |
| $Cl^0$ (2) | 13.64 to 13.68 | 13.61 | 13.62 | 13.62 |
| $Mg^0$ (2) | 13.27 to 13.67 | 13.38 | 13.53 | 13.79 |
| $N^0$ (2) | 16.67 to 17.08 | 17.00 | 17.00 | 17.00 |
| $Fe^0$ (19) | 11.70 to 11.86 | 11.55 | 11.61 | 11.66 |
| $K^0$ (19) | 11.82 to 11.88 | 11.30 | 11.42 | 11.66 |
| $Na^0$ (19) | 13.85 to 13.93 | 13.35 | 13.50 | 13.75 |
| $S^+$ (2) | 16.22 to 16.36 | 16.30 | 16.34 | 16.30 |
| $C^+$ (9) | 17.255 | 17.30 | 17.29 | 17.30 |
| $Si^+$ (2) | 16.45 to 16.82 | 16.70 | 16.70 | 16.70 |
| $N^+$ (2) | 15.75 to 17.20 | 13.75 | 13.74 | 13.74 |
| $Ca^+$ (19) | 11.94 to 11.98 | 12.02 | 12.10 | 12.19 |
| $Cl^+$ (2) | ≤ 13.17 | 11.76 | 11.77 | 11.83 |
| $Mg^{+}$ (2) | 15.71 to 15.74 | 15.60 | 15.60 | 15.59 |
| $Fe^{+}$ (2) | 14.10 to 14.40 | 14.20 | 14.20 | 14.20 |
| $Cu^{+}$ (2) | 11.96 to 12.06 | 12.10 | 12.09 | 12.09 |
| $Mn^{+}$ (2) | 13.42 to 13.70 | 13.55 | 13.56 | 13.56 |
| $S^{++}$ (2) | ≤ 13.55 | 13.25 | 13.19 | 13.19 |
| $H_3^{+}$ (10) | 13.9 | 13.24[a], 12.87[b] | 13.89[a], 13.55[b] | 14.08[a], 13.73[b] |
| NH (11) | 11.85 to 12.05 | 10.91 | 11.43 | 11.77 |
| CO (2) | 14.69 to 15.03 | 13.28 | 14.00 | 14.48, 14.1[c], 14.66[d] |
| OH (18) | 13.57 to 13.65 | 13.34 | 13.97 | 14.45, 14.56[c], 14.0[e] |
| CH (12) | 13.28 to 13.3 | 12.07 | 11.98 | 11.77, 13.42[c] |
| $CH^{+}$ (13) | 12.54 | 9.84 | 9.79 | 9.74, 12.02[c] |
| CN (14) | 12.43 to 12.52 | 9.33 | 9.80 | 9.99, 10.25[c] |
| $C_2$ (15) | 13.20 to 13.34 | 11.05 | 10.71 | 10.01, 11.38[c] |
| $C_3$ (16) | 12 | 7.95 | 7.65 | 6.92, 8.30[c] |
| $H_2(J=0)$ (2) | 20.34 to 20.68 | 20.73 | 20.70 | 20.58 |
| $H_2(J=1)$ (2) | 20.01 to 20.36 | 20.33 | 20.34 | 20.37 |
| $H_2(J=2)$ (2) | 18.02 to 18.08 | 17.56 | 17.64 | 17.84 |
| $H_2(J=3)$ (2) | 16.11 to 16.43 | 15.55 | 15.67 | 15.95 |
| $H_2(J=4)$ (2) | 14.95 to 15.07 | 14.45 | 14.53 | 14.76 |
| $H_2(J=5)$ (2) | 14.32 to 14.40 | 14.06 | 14.21 | 14.50 |
| $T_{10}$ (K) | 46 to 79 | ~ 54 | ~ 56 | ~63 |

[1]Bohlin (1975) [2]Snow (1977) [3]White (1973) [4]van Dishoeck & Black (1986) [5]Chaffee (1974) [6]Hobbs (1974) [7]Hobbs (1969) [8]Chaffee & Lutz (1977) [9]Cardelli et al. (1996) [10]McCall et al. (2003) [11]Meyer & Roth (1991) [12]Jura & Meyer (1985) [13]Federman (1982) [14]Meyer & Jura (1985) [15]van Dishoeck & Black (1989) [16]Maier et al. (2001) [17]Jenkins, E. B., & Shaya, E. J. (1979) [18]Felenbok & Roueff (1996)[19]http://astro.uchicago.edu/home/web/welty/coldens.html [20]Cartledge et al. (2001)[21]Welty et al. (1991)

[a]Using recombination rate of $H_3^{+}$ measured by MC03 [b]Using recombination rate of $H_3^{+}$ measured by Datz et al. [c]Using non-thermal chemistry [d]Using 3/2 times decreased photodissociation rate [e]Using 3 times increased photodissociation rate

**Table 3. Effect of the grain size distribution on column densities with $\chi_{CR}$ = 40**

| Species | Column densities with larger grains (Orion type, *R*v = 5.5 ) | Column densities with smaller grains (ISM type, *R*v = 3.1) | Observed column densities for ζ Per |
|---|---|---|---|
| $H_3^+$ | 12.83 | 13.50 | 14.08 |
| $H_2$ | 20.36 | 20.66 | 20.78 |
| $H_2^+$ | 12.88 | 13.18 | 13.34 |
| $S^0$ | 13.55 | 13.65 | 13.59 |
| $S^+$ | 16.30 | 16.30 | 16.30 |
| $C^0$ | 15.57 | 15.70 | 15.60 |
| $C^+$ | 17.30 | 17.30 | 17.30 |
| Fe | 11.69 | 11.63 | 11.66 |
| $Fe^+$ | 14.20 | 14.20 | 14.20 |

**Table 4: Predicted column densities of undetected species for the model with $\chi_{CR}$ = 10 and 40**

| Species | Predicted log $N$ cm$^{-2}$ ($\chi_{CR}$ = 10) | Predicted log $N$ cm$^{-2}$($\chi_{CR}$ = 40) |
|---|---|---|
| $H^+$ | 18.10 | 18.19 |
| $H_2^+$ | 12.84 | 13.38 |
| $OH^+$ | 12.16 | 12.79 |
| $H_2O^+$ | 12.03 | 12.54 |
| $H_3O^+$ | 12.01 | 12.22 |